# Nucleation-growth versus spinodal decomposition in Fe-Cr alloys: an experimental verification by atom probe tomography and small angle neutron scattering

Sudip Kumar Sarkar [1*,3], Debes Ray[2], Debasis Sen[2,3], Aniruddha Biswas[1,3]

*[1]Materials Science Division, [2]Solid State Physics Division*

*Bhabha Atomic Research Centre, Mumbai-400085*

*[3]Homi Bhabha National Institute, Mumbai-400094*

[*]E-mail: s.sudip.iitg@gmail.com, sudips@barc.gov.in

**Abstract:**

Identifying the operative mode of phase separation (spinodal decomposition (SD) or nucleation-growth (NG)) remains a largely unexplored area of research in spite of its importance. The present work examines this critically in Fe-Cr system using atom probe tomography (APT) and small angle neutron scattering (SANS), and establishes the framework to distinguish the two different modes of $\alpha^{/}$ phase separation in thermally aged Fe-35 at.% Cr and Fe-20 at.% Cr alloys. Independent APT analysis determines the mode of phase separation on the basis of: (i) presence / absence of periodic chemical fluctuation through radial distribution function analysis; and (ii) inter-phase interface characteristics (diffuse / sharp). SANS analysis, in contrast, yields virtually indistinguishable correlation peaks for both the modes, which necessitates further investigation of the several different aspects of SANS profiles in the light of APT results. For the first time, key features of SANS profiles have been identified that can unambiguously distinguish SD from NG in Fe-Cr system: (i) nature of temporal evolution of FWHM of the correlation peak; and (ii) appropriate value of $\gamma$ for fitting with the dynamic scaling model ($\gamma = 6$ for SD, Fe-35 at.% Cr alloy; $\gamma = 4$ for NG, Fe-20 at.% Cr alloy).

1. Introduction

Material properties, be it structural or functional, are often governed by phase separation and the presence of second phase decomposition product in the microstructure. In particular, the characteristics of the second phase and the kinetics of decomposition play crucial roles, both of which are likely to be influenced by the operative mode of phase separation. Understandably, general response of material to nucleation and growth (NG) mode of phase separation is different in comparison to its response to spinodal decomposition (SD). Naturally, there is strong interest among the researchers, especially in the metallurgy community, in experimentally determining the mechanism of phase separation.

Extensive literature and multiple review articles are available on different aspects of the process of phase separation (Wagner et al., 2001; Binder & Fratzl, 2005). In general, a phase separation is initiated when a homogenized supersaturated solid solution is either cooled slowly into the two phase region or quenched into the two phase region prior to thermal aging at a temperature within the two phase region. Typically, depending on the nature of quenching into the miscibility gap in the phase diagram (shallow quench into the metastable region or deep quench into the unstable region), one of the following two modes of phase separation is expected: NG or SD (Aaronson et al., 2016). NG mode is based on heterophase fluctuations and the classical NG model entails formation of precipitates with a fixed composition and a sharp inter-phase interface that delineates it from the matrix (Aaronson, 1970). In contrast, SD (Cahn, 1961; Cahn & Hilliard, 1971) is initiated through homophase concentration fluctuations having diffuse interface, which culminates into sharp inter-phase interface on prolonged aging. Since the transformation products at the final states of NG and SD are microstructurally indistinguishable, studying the early stages of decomposition is critically important in identifying the mode of phase separation (Müller et al., 2015). In addition, it is necessary to monitor the progress of phase separation by interrupting it at different stages, which lets one reconstruct the pathway through snapshots of the interrupted microstructures. In order to do that, micro analytical tools are required that are capable of measuring the (i) spatial extent and amplitude of composition fluctuation for SD; and/or (ii) size, shape, morphology, number density and composition of precipitates for NG mode of phase separation. This effectively restricts the choice of available micro analytical tools (direct imaging and scattering techniques) only to a handful that have the necessary resolution to characterize atomic clusters of a few nanometers size and the capability to determine their chemical composition. Wagner et al. (Wagner et al., 2001) have reviewed the candidate techniques along with their relative merits and limitations: transmission electron microscopy (TEM), field ion microscopy (FIM) and atom probe tomography (APT) for direct imaging, and small angle x-ray scattering (SAXS) and small angle neutron scattering (SANS) among the scattering techniques. It emerges that no single technique can provide a definitive answer. It is prudent to combine different complementary techniques that can potentially determine the mode of phase separation unambiguously (Wagner et al., 2001). This has recently been demonstrated successfully for Cu-34 at.% Ta and Fe-35 at.% Cr alloys using TEM-APT (Müller et al., 2015) and APT-SANS (Sarkar et al., 2021a) combinations, respectively. It may also be noted that very often an interconnected morphology of the decomposed structure is interpreted as SD (Zhou et al., 2013) while discrete precipitates as the mark of NG (Novy et al., 2009). In reality, this approach can be misleading, because NG mode too can give rise to very similar modulated morphology (Ardell, 1967; Doi & Miyazaki, 1986; Doi et al., 1984) or SD might manifest itself as discrete precipitates (Piller et al., 1984). It is, therefore, important to experimentally establish the characteristic distinctive features of the different modes of phase separation. For example, as noted by Muller et al. (Müller et al., 2015), the following two experimentally verifiable attributes can unambiguously prove the spinodal nature of phase separation: (i) increase in amplitude of composition fluctuation; and (ii) concomitant reduction in the width of inter-phase interface, both with increasing aging time. The presence of periodic compositional fluctuation, evidenced by radial distribution function (RDF) analysis of APT data, is another definitive indication of SD (Zhou et al., 2013). On the contrary, no such periodic fluctuation is expected from such RDF analysis in case of NG mode. As for the evolution of width of the inter-phase interface, it is not expected to change in case of classical NG mode. However, there can be deviation in case of non-classical NG, where precipitates are not known to form with a fixed composition (Novy et al., 2009).

In case of binary Fe-Cr alloys, a system with a wide miscibility gap and an ideal one for phase separation study, there is ongoing debate on the position of the boundary line in the phase diagram that separates NG from SD (Andersson & Sundman, 1987). Nevertheless, it is well known that these alloys undergo decomposition into a nano-scale $\alpha\prime$ (Cr-rich) + $\alpha$ (Fe-rich) two-phase microstructure in the temperature region of 573 to 823 K (Danoix et al., 1992; Brenner et al., 1982; Xiong et al., 2010). As regards the characterization of phase separation in this system, APT and SANS are the two most effective techniques and therefore, used extensively (Tissot et al., 2019; Xu et al., 2016). Multiple experimental studies have reported SD mode of phase separation for Cr concentration above 30 at.% in the temperature region of 673–823 K (Brenner et al., 1982; Park et al., 1986; Xiong et al., 2010; Yan et al., 2017). Interestingly, Pareige et al. (Pareige et al., 2011) reported SD at 773 K in an alloy containing much less, 25 at.% Cr. In contrast, NG is reported in binary Fe-Cr alloys containing 19 (Tissot et al., 2019), 20 (Novy et al., 2009) and 25 at.% Cr at 773 K (Xu et al., 2016). However, there is very little work in the literature that focuses on experimental confirmation of the operative mode of phase separation in Fe-Cr system. Barring a few exceptions ( Zhou et al., 2013; Sarkar et al., 2021a), the vast majority of the studies based on APT or APT-SANS combination have chosen morphology of the transformation product as an indicator of the underlying mode of phase separation in this system. In some cases, the authors have accepted *a priori* a certain mode of phase separation to be the operative one for their alloy (Pareige et al., 2011). As for the independent SANS analysis in Fe-Cr system, there is additional complication that needs to be discussed in further detail. For example, even though it is a well known fact that appearance of a correlation peak (interference peak) in SANS pattern is associated with the periodic compositional fluctuation of SD, it is not necessarily a unique and exclusive attribute of only the SD mode of phase separation. This is elaborated later in this section. However and not surprisingly, most of the SANS work on SD in Fe-Cr alloys have focused on studying the change of structure factor by measuring the scattered SANS intensity and simultaneously comparing with either Cahn-Hilliard Cook (CHC) theory or Langer-Bar-on-Miller (LBM) theory (Bley, 1992; Furusaka et al., 1983; LaSalle & Schwartz, 1986). The SD process may be described as the change of spatial distribution of the alloy component, which can be represented by a nonlinear evolution equation based on the diffusion equation. The experimental scattered intensity is proportional to the theoretical structure factor that is nothing but the Fourier transform of the two-body correlation function of local composition. As alluded to earlier, similar correlation peak is also observed in quite a few cases of NG mode of phase separation in Fe-Cr system. For example, though Fe-20 at.% Cr alloy is expected to follow NG as per the APT and MS studies (de Nys & Gielen, 1971; Novy et al., 2009), it still shows correlation peak in SANS pattern (Tissot et al., 2019). Similar correlation peak has also been observed in other alloy systems as well, such as $\delta^{/}$ precipitate in Al- 9.7Li alloy (Tsao et al., 1999) and α$^{/}$ precipitate in Fe-Cr-Al (Briggs et al., 2017) alloys, both of which are interpreted as NG type precipitates. Therefore, as mentioned before, the presence of an interference peak (correlation peak) in SANS pattern cannot be considered as a unique characteristic of SD. The work by Bley et al. is particularly relevant in this regard (Bley, 1992). They have studied the un-mixing kinetics using SANS in binary Fe-Cr alloys containing 20, 35 and 50 at.% Cr (Bley, 1992). Their results show no difference between the nature of the SANS patterns of 20 at.% Cr (NG) and 35/50 at.% Cr alloys (SD). In both the cases, the correlation peak appears in the SANS pattern and both the cases are consistent with the CHC-LBM kinetics. Therefore, it is necessary to look closely at the capabilities of SANS in distinguishing NG and SD, which has not been explored so far in the literature.

With this background, in this work, we have critically examined how effective APT and SANS are in unambiguously distinguishing the two modes of phase separation namely, NG and SD in Fe-Cr alloy. Binary Fe-20 at.% Cr is the primary alloy of choice, which is generally accepted as the one to undergo NG mode of phase separation on aging at 773 K (Chandra & Schwartz, 1971). The result of Fe-20 at.% Cr is compared with that of Fe-35 at.% Cr alloy, which has already been experimentally verified to undergo SD mode of phase separation (Sarkar et al., 2021a). Therefore, the current article has two major parts. The first part focuses on experimental confirmation of NG mode of phase separation in Fe-20 at.% Cr alloy by APT, which is not available in the literature. This is achieved on the basis of RDF analysis and temporal evolution of the inter-phase interface. In the second part, special emphasis is given on identifying the footprints from SANS analysis that can serve as a potential indicator of the underlying mechanism of phase separation. It scrutinizes several key aspects of the experimental SANS patterns of both Fe-20 at.% Cr and Fe-35 at.% Cr alloys, like evolution of Porod's exponent and full width half maximum (FWHM) of the correlation peak and the fitting characteristics in terms of the dynamic scaling model of SD (Furukawa, 1984). This exercise successfully demonstrates that SANS analysis too is capable to conclusively differentiate one mode from the other in Fe-Cr system.

## 2. Experimental details

The buttons of binary Fe-20 at.% Cr alloy were prepared by vacuum arc melting high purity elements in appropriate proportions in water cooled copper hearth using tungsten electrode. The as-synthesized buttons were re-melted multiple times to ensure homogeneity. The arc-melted alloy was subsequently hot-rolled to reduce thickness. The rolled sample was sealed in a quartz ampoule filled with helium gas, solutionized at 1273 K for 24 h, and quenched in ice water. Samples of dimension 10 mm x 10 mm x 1 mm were cut from the rolled strip by electro-discharge machining and subjected to thermal aging at 773 K for the following durations: 1, 25, 50, 120, 240, 500, 750 and 1000 h. This heat treatment was followed by ice quenching. Detailed characterization of the solutionized and aged samples has been carried out using x-ray diffraction (XRD), x-ray fluorescence with WDS detector (XRF), APT and SANS. XRD experiments carried out using Cu $K_\alpha$ radiation at room temperature confirmed the presence of bcc ferrite phase in the solutionized samples (both Fe-20 at.% Cr and Fe-35 at.% Cr alloys). The bulk chemical composition of Fe-20 at.% Cr alloy as obtained from XRF analysis is presented in Table 1. For the sake of comparison, compositional analysis of Fe-35 at.% Cr alloy (from (Sarkar et al., 2021a)) is also included in the table. Alloy compositions are found to be close to the nominal compositions.

Specimens for FIM and APT are prepared by standard 2-stage electro-polishing of blanks of dimension 0.3 mm x 0.3 mm x 8 mm. Coarse polishing is carried out with 10 vol. % perchloric acid in glacial acetic acid followed by fine polishing with 2 vol. % perchloric acid in 2-butoxyethanol. FIM imaging has been performed with Ne at a pressure of 2 x $10^{-5}$ Torr. Both FIM and APT experiments are performed at 30 ± 0.5 K using Cameca$^{TM}$ FlexTAP. APT experiments are performed in laser-pulsing mode with femto-second UV (343 nm) laser pulse energy of 30 nJ and pulse repetition rate of 50 kHz. No experiment is carried out in voltage pulsing mode as FlexTAP is not equipped with one. Diaphragm has been set at 15-degree configuration for better mass resolution. Data analysis is carried out using IVAS 3.8.0 software.

Adequate care has been taken for the reconstruction of APT data, the details of which is reported in a recent article by the same authors (Sarkar et al., 2021a). FIM analysis of the specimen prior to APT analysis has helped in deducing the image compression factor (ICF). Additionally, the reconstruction parameters have been calibrated in terms of the inter-planar spacing of (110) planes that are identified in the atom map. Satisfactory reconstruction has been achieved by using a field factor value of 6.7 and evaporation field of (29 V/nm). The same reconstruction parameters are followed for all the aged samples. Average core compositions of the precipitates and that of the matrix have been determined from proximity histogram analysis. An average of five data points at the centre of the precipitate denotes the precipitate core composition. For the matrix composition, an average of ten data points far from the interface has been used. All error bars presented correspond to the standard deviations from the mean for the compositional analyses.

In order to investigate the mesoscopic structure of the alloys, SANS and medium resolution SANS experiments have been performed on the aged alloys accessing a wide range of wave vector transfer ($q$), using two facilities: i) SANS-I (Aswal & Goyal, 2000) and ii) SANS-II (MSANS) (Mazumder et al., 2001) facilities, respectively at Dhruva Reactor, Mumbai. In MSANS, the scattering at lower '$q$' domain (i.e. 0.003-0.173 nm$^{-1}$) has been performed using neutron beam of wavelength ($\lambda$) 0.312 nm ($\Delta\lambda/\lambda$ ~ 1%) using double crystal (Si[111]) in non-dispersive geometry. In SANS-I, the mean wavelength of the monochromatized beam from neutron velocity selector is 0.52 nm with a spread of $\Delta\lambda/\lambda$ ~ 15%. The angular distribution of neutrons scattered by the sample is recorded using a 1 m long one-dimensional He$^3$ position sensitive detector. The instrument covers a $q$-range of 0.15–2.5 nm$^{-1}$. The SANS data have been corrected for background and transmission. Further, MSANS data have been de-convoluted to correct for the instrument resolution prior to further analysis. Data from SANS-I and MSANS facilities have been normalized at common $q$ region (0.17 nm$^{-1}$) of the two data sets, where the functionality of the profiles from both the facilities remains more or less similar, in order to get a continuous profile over the total accessible $q$ range.

3. Results

3.1. APT Analysis

*3.1.1. Morphological evolution of* α$^/$ *phase*

The distribution of Cr atoms for the aged (50 h to 1000 h) specimens is displayed in Figure 1a-f. Each purple dot represents one Cr atom. A reconstruction slice of cross-sectional dimension 30 nm × 30 nm with a thickness 10 nm is presented for each aging condition for better view of the temporal evolution of microstructure. Progression of phase separation is clearly visible from these comparative atom maps that show the evolution of Cr rich α$^/$ phase, interrupted at different aging times. They start becoming noticeable 50 h onwards. Therefore, atom maps for the solutionized, 1 h and 25 h aged samples are not presented. The morphology of these Cr rich α$^/$ precipitates can be reasonably approximated as spherical. They are found to be spatially isolated and discrete in nature. The size of the precipitates appears to increase with aging time while their number density declines. Unlike the typical SD mode of phase separation in Fe-Cr system (Yan et al., 2017; Sarkar et al., 2021a), no interconnected network of the Cr-rich second phase is noticed in this case.

### 3.1.2. *RDF analysis of APT data: estimation of periodic compositional fluctuations*

RDF analysis assesses the average local neighborhood as a function of distance extending radially outwards from each atom in the APT dataset and the average radial concentration profiles for all the detected atoms are calculated. It can be described as the probability density of finding an atom *j* at r (x, y, z) when an atom *i* is at origin (Miller & Kenik, 2004; Geuser et al., 2006). Step size of 0.1 nm is employed for the RDF analysis and the measured concentration at each position has been normalized with the average bulk composition. Mathematically, RDF is written as (Zhou et al., 2012, 2013):

$$RDF = \frac{C_E(r)}{C_0} = \frac{N_E(r)/N(r)}{C_0} \quad \ldots\ldots(1)$$

where $C_E(r)$ is the atomic composition of element E at the distance of *r*, $C_0$ is the average composition of element E in the analyzed volume, $N_E(r)$ is the total number of all atoms at the distance of *r*.

As an example, the bulk normalized concentration (Cr-Cr) profile obtained from the RDF analysis for the 1000 h aged specimen of Fe-20 at.% Cr alloy is shown in Figure 2a. The inset magnifies the most relevant portion of the same plot above a Z (radial distance) value of 4 nm. The inset also contains the results from a similar analysis for a sample of Fe-35 at.% Cr alloy aged at 773 K for 500 h (Sarkar et al., 2021a), which shows a very clear peak (first maxima) at a distance of 7 nm. This peak basically corresponds to the statistically averaged distance between two Cr rich regions in a 3-dimensional dataset and represents the wavelength of SD mode of phase separation (Zhou et al., 2013). In contrast, no such first maxima is detected in the RDF analysis of Fe-20 at.% Cr alloy in the current investigation. It, thus, rules out the presence of any 3-dimensional periodic compositional fluctuations in Fe-20 at.% Cr alloy and indicates spatially random distribution of $\alpha^{/}$ precipitates.

### 3.1.3. Probing the temporal evolution of α-α$^{/}$ interface width using APT-based analysis

The nature of temporal evolution of the width of α-α$^{/}$ inter-phase interface can potentially be a key to understanding the operative mechanism of phase separation in Fe-Cr system. For that, Cr concentration profile along the distance (Z) from the interface as obtained from the proximity histogram analysis is utilized, Figure 2b. It is possible to estimate the width of an interface ($\delta$) from the concentration profile by using the following relationship (Ardell, 2012):

$$\delta = \frac{\Delta X}{\left(\frac{dX}{dZ}\right)_{max}} \ldots\ldots\ldots\ldots\ldots\ldots\ldots\ldots\ldots\ldots\ldots\ldots(2)$$

where $\Delta X$ is the the composition difference between the phases and $\left(\frac{dX}{dZ}\right)_{max}$ is the highest slope of the interface concentration profile. $\left(\frac{dX}{dZ}\right)_{max}$ is calculated from the peak value of Gaussian fitted $\frac{dX}{dZ}$ plot, as shown in the inset of Figure 2b. It should be pointed out that this relationship is valid only for coherent interface. The temporal evolution of the as-calculated interface width is presented in Figure 2c along with the result for Fe-35 at.% Cr alloy (Sarkar et al., 2021a). It becomes immediately apparent that there is a striking difference between the natures of temporal evolution of the interfaces of Fe-20 at.% Cr and Fe-35 at.% Cr alloys. The change in the interface width with increasing aging time for Fe-20 at.% Cr alloy appears insignificant when compared with that for Fe-35 at.% Cr alloy.

Based on the following two key results: (i) absence of first maxima in the RDF plot; (ii) virtually flat temporal evolution of α-α$^/$ inter-phase interface width, it is clear that Fe-20 at.% Cr alloy follows NG mode of phase separation. Subsequently, quantification of these α$^/$ precipitates are carried out using a combination of APT and SANS for further kinetic analysis.

### 3.2. Quantification of α$^/$ precipitates using APT and SANS

Isolated nature of the Cr-rich α$^/$ is amenable to delineation from the matrix by appropriate Cr iso-concentration surfaces. Suitable Cr threshold value (35 at.%) for the iso-concentration surface is deduced on the basis of the size information derived from the TEM analysis, details of which is reported in an earlier work by the authors (Sarkar et al.; 2021b). It should be pointed out that quantification can also be carried out by independent APT analysis using maximum separation based cluster search algorithm. However, our earlier study has demonstrated that the iso-concentration based technique following a combinatorial method using APT, TEM and SANS provides more reliable quantification in this case. Figure 3 shows an example of one such reconstructed volume for the 240 h aged specimen where α$^/$ precipitates are isolated from the matrix using 35 at.% Cr iso-concentration surfaces. APT-based detailed quantification of the average precipitate size, number density and volume fraction of the α$^/$ precipitates is performed afterwards, results of which are presented later along with the SANS-derived ones. Compositional information derived from the APT analysis is used subsequently to determine the scattering length contrast for SANS profiles, Figure 4. These profiles can be divided primarily into three zones. Lower $q$ region (Zone-I, $q < 0.05$ nm$^{-1}$) corresponds to large structures like magnetic domains (Xu et al., 2019), intermediate $q$ range (Zone-II, $0.05 < q < 0.2$ nm$^{-1}$) corresponds to some intermediate size structures while signal coming from q region greater than 0.2 nm$^{-1}$ (Zone-III) signifies structures arising from phase separation. In order to study the Cr-rich phase separation of spatial dimension of a few nm, data above $q$ region greater than 0.2 nm$^{-1}$ (Zone-III) is analyzed rigorously. The spherical equivalent radius, number density and volume fraction of α$^/$ precipitates for different aging durations that are obtained from APT and SANS analyses are shown together in Figure 5a-c. These results show excellent agreement between the two analysis techniques. The precipitate size shows gradual increase with aging time with concomitant decline in the number density. The volume fraction does not saturate even after 1000 h of aging, Figure 5c. Temporal evolution of the phase compositions is presented in Figure 5d (from APT analysis only). Precipitate size vs time plot follows $\sim t^{1/3}$ relationship (inset of Fig. 5a), which is in line with LSW theory of coarsening (Lifshitz & Slyozov, 1961; Wagner, 1961). An increase in precipitate size, decrease in number density, non-attaining of constant volume fraction suggest transient coarsening, which is the superposition of nucleation with coarsening processes (Wagner et al., 2001; Novy et al., 2009).

Once it is established by APT analysis that it is the NG mode that is operative in Fe-20 at.% Cr alloy, the next step is to look for any similar potential indicator in the SANS profiles. That way, SANS analysis too would be able to identify the actual mode of phase separation in Fe-Cr system. Accordingly, the following section examines different aspects of the SANS profiles with that aim to distinguish the two modes of phase separation, i.e., NG and SD.

### 3.3. SANS analysis: deciphering the mode of phase separation

A closer look at the SANS profiles for Fe-20 at.% Cr alloy in Figure 4 confirms the presence of a correlation peak at around $q \sim 0.7$ nm$^{-1}$, which can be attributed to either a wavelength modulation of the chemically spinodal structures (Furusaka et al., 1983; LaSalle & Schwartz, 1986) or some kind of a correlation among the densely populated NG precipitates (Tsao et al., 1999; Briggs et al., 2017; Tissot et al., 2019). To unearth the true nature of phase decomposition from the SANS profiles, additional analysis is required.

*3.3.1. Evolution of Porod's exponent*

The slope of log ($I$) vs log ($q$) in a $q$ region beyond the correlation peak is often considered one of the characteristic features that is associated with the chemical nature of the interface of the decomposed structures (Hörnqvist et al., 2015). Therefore, the change in slope, i.e., Porod's exponent, is determined from the plot between $I(q)q^4$ and $q$ for all the samples of Fe-20 at.% Cr alloy, aged for 50 h and more, Fig. 6(a). Solutionized, 1 h and 25 h aged samples do not show any correlation peak in the scattering pattern. An exponent value of 2.5 is noticed for the 50 h aged sample, which changes to 3.6 when aging time becomes 120 h, thereafter a slope of 4 is observed. From 240 h onwards, a clear plateau appears just after the $q_{max}$ ($q$ corresponds to the maximum intensity of the correlation peak) in the $I(q)q^4$ vs $q$ plot. It is demonstrated using the example of the 1000 h aged sample, where the position of $q_{max}$ is indicated by an arrow. This clearly suggests that intensities fall as $q^{-4}$ after $q_{max}$ for samples aged for 240 h and more. Figure 6b compares the temporal evolutions of Porod's exponents of Fe-20 at.% Cr and Fe-35 at.% Cr alloys (Sarkar et al., 2021a). The inset of Figure 6b shows an example of a similar analysis carried out for Fe-1.4 at.% Cu alloy (aged at 773 K), which is known to follow classical NG mode for Cu-rich precipitates (Ahlawat et al., 2019). In that case, Porod's exponent of 4 is observed from aging time as low as 3 h.

*3.3.2. Evolution of the FWHM of the correlation peak*

The second aspect of the SANS profiles that one can analyze is whether the evolution of the shape of the correlation peak reveals any useful information about the nature of phase separation. In order to find out any possible correlation, the FWHM values are computed after fitting the correlation peaks with a Gaussian function for all the pertinent aging conditions. These FWHM values are plotted as a function of aging time for Fe-20 at.% Cr alloy and the same has been compared with the equivalent results from Fe-35 at.% Cr alloys (exhibits SD mode (Sarkar et al., 2021a)), Figure 7a. An obvious difference between them in the evolution of FWHM of the correlation peak is clearly visible. In case of SD, the correlation peaks become sharper more rapidly as the phase separation progresses and as a result, FWHM decreases exponentially with aging time for Fe-35 at.% Cr alloy. As shown later in the Discussions section, this exponentially decaying behaviour is not surprising and can be justified in terms of the theoretical CHC model for SD (Bley, 1992; Hörnqvist et al., 2015), and the same has been demonstrated in Figure 7b for Fe-35 at.% Cr alloy. On the contrary, for Fe-20 at.% Cr alloy, the FWHM values only show a marginal decline in a linear fashion. These comparative results make it abundantly clear that Fe-20 at.% Cr alloy does not follow the peak shape evolution like the spinodally decomposed Fe-35 at.% Cr alloy (Hörnqvist et al., 2015).

*3.3.3. Fitting characteristics with the dynamic scaling model of SD*

Furthermore, correlation peaks for Fe-35 at.% Cr and Fe-20 at.% Cr alloys have been fitted with the dynamic scaling model of SD that was first proposed by Furukawa et al. and used extensively ever since (Furukawa, 1984; Xu et al., 2019, 2020; Sarkar et al., 2021a):

$$I(q) = \frac{I_{max}\,(1+\gamma/2)x^2}{(\gamma/2+x^{2+\gamma})} + B_g \dots\dots\dots(3)$$

Typically, the spinodally decomposed systems show a characteristic interference peak reflecting the correlation length ($\sim \frac{2\pi}{q_{max}}$), where $q_{max}$ denotes the peak position. $I_{max}$ is the scattering intensity at the peak position and $x = q/q_{max}$. $B_g$ is the background contribution. The exponent ($\gamma$) is determined from the dimensionality of the system ($D$) (Furukawa, 1984). For example, its value is equal to $2D$ for critical concentration mixtures above the percolation threshold, which is the case for SD mode of phase separation. Therefore, $\gamma = 6$ is expected for a 3-dimensional analysis. Otherwise, $\gamma$ is equal to $(D + 1)$ for the system below the percolation threshold. In case of random discrete 2-phase mixture like that of a NG-derived one, $\gamma = 4$ will be the appropriate one. The characteristic correlation peaks along with the best fittings for both Fe-35 at.% Cr and Fe-20 at.% Cr alloys are presented in Figure 8a,b.

As expected, all the SANS profiles for Fe-35 at.% Cr alloy can be fitted well with the spinodal model if a $\gamma$ value of 6 is considered, Figure 8a. In contrast, in case of Fe-20 at.% Cr alloy, they can only be fitted with a the SD model when a $\gamma$ value of 4 is used, Figure 8b. Using an example of the 1000 h aged sample of Fe-20 at.% Cr alloy, inset of Figure 8b compares the fittings with the SD model for a $\gamma$ value of both 4 and 6. It is evident that the quality of fitting is unacceptably poor with a $\gamma$ value of 6, as opposed to a $\gamma$ value of 4 ($R^2$ value of 0.99 and 0.74 for $\gamma$ value of 4 and 6, respectively). The result suggests that Fe-20 at.% Cr alloy is well below the percolation threshold, which in turn, implies that it doesn't follow the SD mode of phase separation.

*3.3.4. Kinetics analysis using SD model*

Lastly, the kinetics associated with the shifting of the correlation peak is investigated in light of the dynamic scaling model of SD in order to find out any possible relationship between the dynamics of decomposition and the mode of phase separation. The time evolution of scattering profiles is characterized by shifting of $q_{max}$ and $I_{max}$ (intensity at $q_{max}$). In case of SD, both $q_{max}$ and $I_{max}$ are known to follow power law variations with time (Binder & Stauffer, 1974; Marro et al., 1975, 1979):

$$q_{max} \propto t^{-a'} \ldots\ldots\ldots\ldots(4)$$

$$I_{max} \propto t^{a''} \ldots\ldots\ldots\ldots(5)$$

There are a few reports that suggest $a'' = 3a'$ for systems undergoing SD (Binder & Stauffer, 1974; Tsao et al., 1999). As for the values of these exponents for SD, theoretical prediction by Binder et al. (Binder & Stauffer, 1974) suggests the following: $a' = 1/6$ and $a'' = 1/2$. Slightly different values are predicted for SD by Marro et al. (Marro et al., 1975, 1979): $a' = 0.2 - 0.28$ and $a'' = 0.65 - 0.74$. Interestingly, experimental results have demonstrated that these predictions often do not hold good (Katano & Iizumi, 1983). Table 3 lists several such examples from Fe-Cr system that deviate from the predictions. Nevertheless, necessary analysis is carried out to ascertain this point. Accordingly, the values of $a'$ and $a''$ are determined from log-log plots of $q_{max}$ and $I_{max}$ with time (Fig. 9a,b). For Fe-20 at.% Cr alloy, the values are $0.2 \pm 0.01$ and $0.87 \pm 0.02$, respectively. In comparison, $a'$ value for Fe-35 at.% Cr alloy is $0.15 \pm 0.01$ for initial 120 h of aging that changes later to $0.29 \pm 0.02$, while $a''$ value is found to be $1.13 \pm 0.2$. In the current case too, the values of $a'$ and $a''$ turn out to be inconclusive in identifying the mode of phase separation.

## 4. Discussions

Experimental identification of the operative mode of phase separation involves quantitative assessment of the characteristic attributes that are associated with each mode of phase separation.

As explained before, the current work, focuses first on the specific APT-based analyses that help in establishing the mode of phase separation operative in the alloy of interest, thermally aged Fe-20 at.% Cr alloy. The comparative results of RDF analysis of the APT data from the long-duration aged samples of Fe-20 at.% Cr and Fe-35 at.% Cr alloys (Fig. 2a) demonstrate the difference in the nature of one such characteristic attribute namely, periodic compositional fluctuation. Absence of any primary maxima in Fe-20 at.% Cr alloy rules out the presence of any periodic compositional fluctuation, which is a distinctive feature of the SD mode of phase separation. Fe-35 at.% Cr alloy, on the other hand, is known to undergo SD and as expected, shows a very clear primary maxima. Furthermore, in case of Fe-20 at.% Cr alloy, the randomly distributed discrete nature of the Cr-rich $\alpha'$ precipitates (Fig. 1b-h) agrees well with the absence of this peak in the RDF analysis. Another important and experimentally verifiable aspect of phase separation that can be quantitatively estimated by APT analysis is the width of the inter-phase interface and its evolution with aging time (Fig. 6b). It is well known that SD is characterized by chemically diffuse interface at the early stages that gets sharper as the phase separation progresses. This transition of the interface is manifested in Fe-35 at.% Cr alloy by the significant reduction in the interface width with increasing aging time (1.92 nm for 50 h to 1.55 nm for 500 h) (Sarkar et al., 2021a). In contrast, reduction in the interface width for Fe-20 at.% Cr alloy (1.57 nm for 50 h to 1.53 nm for 1000 h) is insignificant and rather negligible. Thus, APT-based analyses have experimentally confirmed the NG mode of phase separation in Fe-20 at.% Cr alloy. At this point it is worthwhile to mention that, the Cr concentration of $\alpha'$ precipitates have continually increased with aging even after 1000 h, while the matrix has gotten increasingly deficient in Cr (Fig. 5d). The temporal evolution of Cr concentration profile is in contradiction to the classical theory of nucleation and indicates the occurrence of non-classical nucleation in Fe-20 at.% Cr alloy (Wagner et al., 2001). As intended, having established the presence of non-classical NG mode of phase separation in Fe-20 at.% Cr alloy using APT, the next part of the work delves into the efficacies of SANS analysis in determining the operative mode of phase separation in Fe-Cr system. This is particularly important because SANS, unlike APT, investigates much larger volume of material (~ 10 -100 mm$^3$) and provides statistically averaged bulk information ( Xu et al., 2016; Tissot et al., 2019).

In order to distinguish between the modes of phase separation, the very first thing that is analyzed in the SANS profiles for any possible indication is the interface character that gets reflected in Porod's exponent (Fig. 6 a,b). In general, a diffuse interface is associated with a Porod's exponent close to 2, while a value of 4 corresponds to a sharp interface ( Guinier et al., 1955; Hörnqvist et al., 2015). 120 h onwards, sign of a sharp interface becomes quite evident in case of Fe-20 at.% Cr alloy, the exponent value rises first to 3.7, quickly reaches 4 and remains constant thereafter. But, the evolution of Porod's exponent does not appear to be a credible indicator. Firstly, very similar nature of evolution of Porod's exponent is also noticed in case of the spinodally decomposed Fe-35 at.% Cr alloy (Fig. 6b). Moreover, the value of Porod's exponent for the early stage (i.e., 50 h of aging) of phase separation in Fe-20 at.% Cr alloy is far from 4. Earliest perceptible evidence of phase separation in terms of a correlation peak in Fe-20 at.% Cr alloy is observed after 50 h of aging, which shows a Porod's exponent value of 2.5. One of the possible reasons for this value of 2.5 could be the non-classical nature of nucleation of $\alpha^/$ (Novy et al., 2009; Hatzoglou et al., 2019). This value is not commensurate to the NG mode of phase separation that entails presence of a sharp interface with an expected exponent value close to 4. An example of NG-type evolution of Porod's exponent is shown in the inset of Figure 8a for Fe-1.4 at.% Cu alloy where a constant value of 4 is attained quickly. In fact, as noted earlier, this value of 2.5 for the 50 h aged sample of Fe-20 at.% Cr alloy is very close to the one that is expected typically for a chemically diffuse interface (a value of 2) (Hörnqvist et al., 2015). For example, at the same initial stage of phase separation (50 h at 773 K) in SD mode in Fe-35 at.% Cr alloy, a diffuse interface with an exponent of 2 is recorded, Figure 6b (Sarkar et al., 2021a). It turns out that Porod's exponent, unlike APT, is not sensitive enough to distinguish between the early stages of NG and SD modes of phase separation in this system. APT analysis, however, has revealed the significant difference between their interface widths at that stage; the APT-derived inter-phase interface width after 50 h of aging for Fe-20 at.% Cr and Fe-35 at.% Cr alloy is 1.57 nm and 1.92 nm, respectively (Fig. 2c). Therefore, no definitive conclusion can be drawn on the basis of Porod's exponents possibly because they depend on many different parameters (Ujihara & Osamura, 2000).

The next important parameter that is examined in this work is the FWHM of the correlation peaks in the SANS profiles of Fe-20 at.% Cr and Fe-35 at.% Cr alloys and their temporal evolution, Fig. 7(a). FWHM of the correlation peak for Fe-35 at.% Cr alloy decreases exponentially as the phase separation progress, while for Fe-20 at.% Cr alloy, it marginally decreases in a linear fashion. As shown in Figure 7a, FWHM profiles for Fe-35 at.% Cr alloy fit very well ($R^2$ = 0.99) with an exponential decay function, which is consistent with the CHC model of SD. According to CHC model, evolution of structure factor $(S(q,t))$ for spinodally decomposed correlation peak can be written as follows (Bley, 1992):

$$S(q,t) = \left\{S(q,0) - \frac{RT}{c(1-c)(A+2kq^2)}\right\} \times \exp\{-2q^2 Mt(A + 2kq^2)\} + \frac{RT}{c(1-c)(A+2kq^2)} \quad \ldots\ldots\ldots(6)$$

where, $S(q, 0)$ is the structure factor for solutionized sample, $c$ is the average Cr content in the alloy, $A$ is the second derivative of free energy with composition, $k$ is the gradient energy component and *M* is the atomic mobility. Since the second term of this equation (equation (6)) represents an exponential decay function with time, the same functional dependency is being reflected in the temporal evolution of the FWHM of the correlation peak for Fe-35 at.% Cr alloy. In order to validate it further, experimental SANS correlation peaks for the aged samples (50, 120, 240 and 500 h) of Fe-35 at.% Cr alloy are simulated with CHC model as per equation (6). Necessary initial values of the parameters for this simulation are available in the literature (e.g., $S(q, 0) = (1-x_0^2) = 0.8775$ for Fe-35 at.% Cr and *M* is 3.49 x $10^{-26}$ m²-mol/J/s (Hörnqvist et al., 2015), $A \sim -366$ J/mol and $k \sim 196$ J/nm²/mol (Bley, 1992)). All the experimental correlation peaks are successfully simulated and the fitting parameters are listed in Table 2. Therefore, it can be said that the temporal evolution of FWHM of the correlation peak for Fe-35 at.% Cr alloy is governed by the parameters *M*, *A* and *k* (present in the exponential term of equation (6)). In contrast, temporal evolution of FWHM of the correlation peak for Fe-20 at.% Cr alloy does not follow exponential decay behaviour as per equation (6), which suggests that Fe-20 at.% Cr alloy does not follow SD. Thus, FWHM of SANS correlation peak acts as a strong indicator to distinguish one mode of phase separation from the other.

Subsequently, when attempts are made to fit the SANS profiles of Fe-20 at.% Cr alloy using the SD model, it becomes quite clear that it necessitates the use of an unrealistic $\gamma$ value of 4 (Fig. 8b). It should be reiterated that for a SD mode of phase separation in a 3-dimensional alloy system, the value of $\gamma$ should be equal to 6 (i.e., 2*D*, where $D = 3$, the dimensionality of the system) (Furukawa, 1984). This is demonstrated by the SANS profiles of Fe-35 at.% Cr alloy in Fig. 8(a). Likewise, a $\gamma$ value of 6 has also been reported earlier for SD in Fe-53 at.% Cr alloy when aged at 748 K by other researchers (Xu et al., 2020). As shown in Figure 8b (as an inset),

SANS profiles of Fe-20 at.% Cr alloy cannot be fitted with a $\gamma$ value of 6. This result further establishes that SANS can also effectively distinguish NG from SD.

However, the power-law variation ($q_{max}$ and $I_{max}$ with time) exponents $a'$ and $a''$ could not provide any conclusive proof by which SD can be separated from NG.

## 5. Conclusions

The current work utilizes APT-based analysis and successfully resolves one long-standing SANS profile puzzle as both the modes of phase separation (SD and NG) manifest virtually indistinguishable correlation peaks. This is accomplished by identifying two key indicators from SANS profiles that can unambiguously distinguish one mode of phase separation from the other. The nature of temporal evolution of FWHM of the correlation peak is one such indicator. For SD mode of phase separation (e.g., Fe-35 at.% Cr alloy), it conforms to the structure factor evolution as per CHC model. NG mode of phase separation (e.g., Fe-20 at.% Cr alloy) does not follow this. The second key indicator is the value of $\gamma$ appropriate for fitting the SANS profile with the dynamic scaling model. The value of $\gamma$ makes clear distinction between the modes of phase separation: $\gamma = 6$ for SD (e.g., Fe-35 at.% Cr alloy), but $\gamma = 4$ (Fe-20 at.% Cr alloy).

**Acknowledgements:**

The authors would like to acknowledge Dr. V. K. Aswal, Solid State Physics Division, BARC for his help in SANS measurements and Dr. Deodatta Shinde, Materials Science Division, BARC for many fruitful discussions.

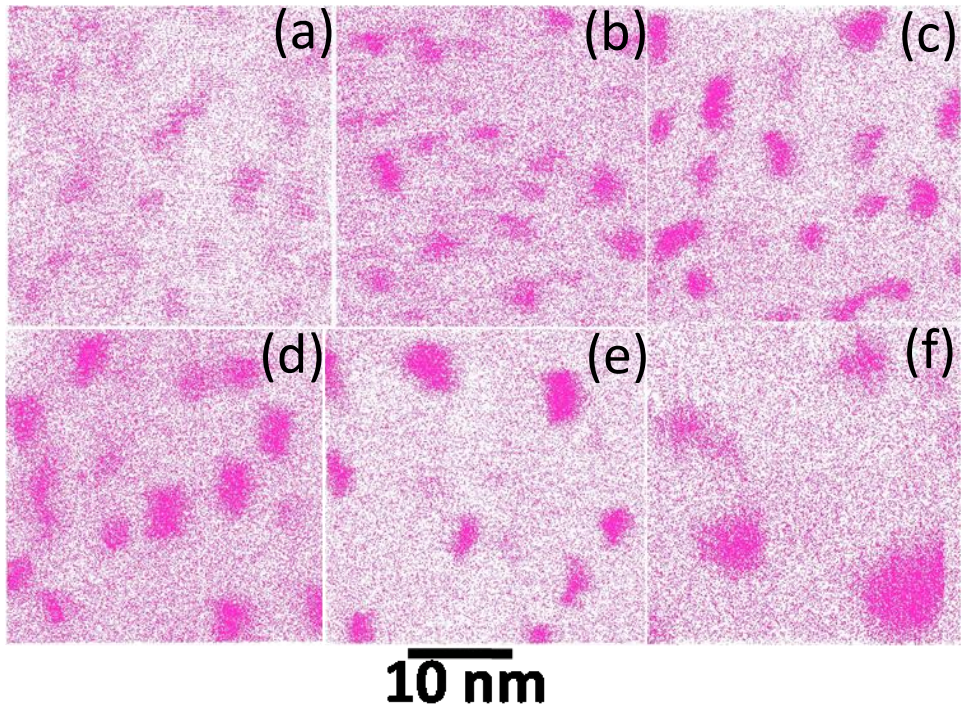

**Fig. 1.** Temporal evolution of α′ as evident from the Cr atom maps of binary Fe-20 at.% Cr alloy aged for: (a) 50 h, (b) 120 h, (c) 240 h, (d) 500 h, (e) 750 h and (f)1000 h respectively. Each purple dot represents one Cr atom.

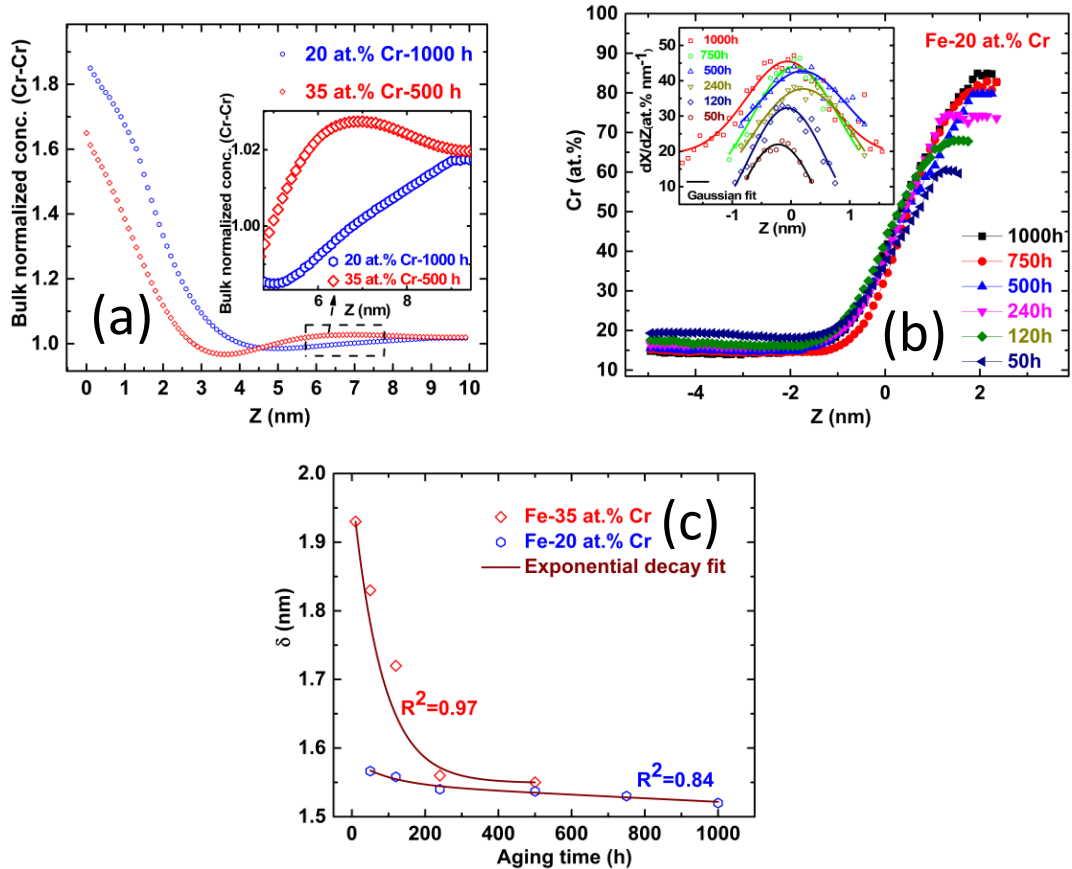

**Fig. 2.** (a) The bulk normalized concentration as obtained from $RDF_{Cr-Cr}$ as a function of distance, shown for 1000 h aged sample for Fe-20 at.% Cr alloy and 500 h aged sample for Fe-35 at.% Cr alloy, for a comparison, while inset shows presence and absence of primary maxima for Fe-35 at.% Cr and Fe-20 at.% Cr alloys respectively when image is magnified for $Z$ values above 5 nm, (b) Proximity histogram profiles displaying the variation of Cr concentration as a function of $Z$ for 50 h onward aged specimens as displayed for Fe-20 at.% Cr alloy; inset shows variations of $dX/dZ$ as function of $Z$, and (c) temporal evolution of interface width for Fe-20 at.% Cr and Fe-35 at.% Cr alloys respectively along with exponential decay fit (solid line).

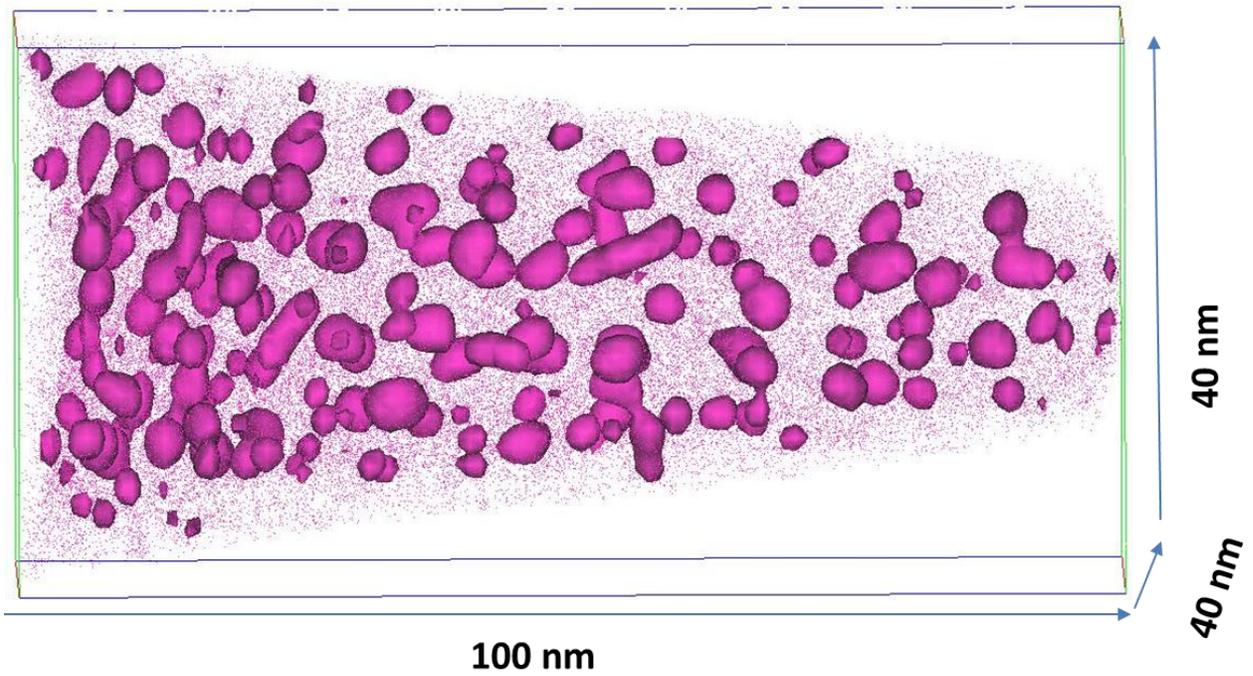

**Fig. 3.** Cr-rich α′ phase separated from matrix using iso-concentration surface with 35 at.% Cr threshold, as shown for 240 h aged Fe-20 at.% Cr alloy sample, displayed as an example.

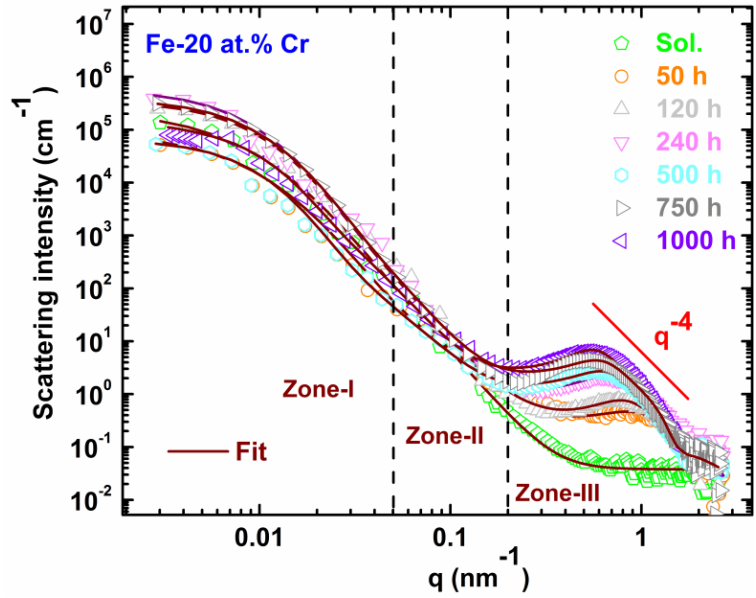

**Fig. 4.** Temporal evolution of SANS pattern (scattered points) for Fe-20 at.% Cr alloy aged at 773 K with spherical model fit (solid line).

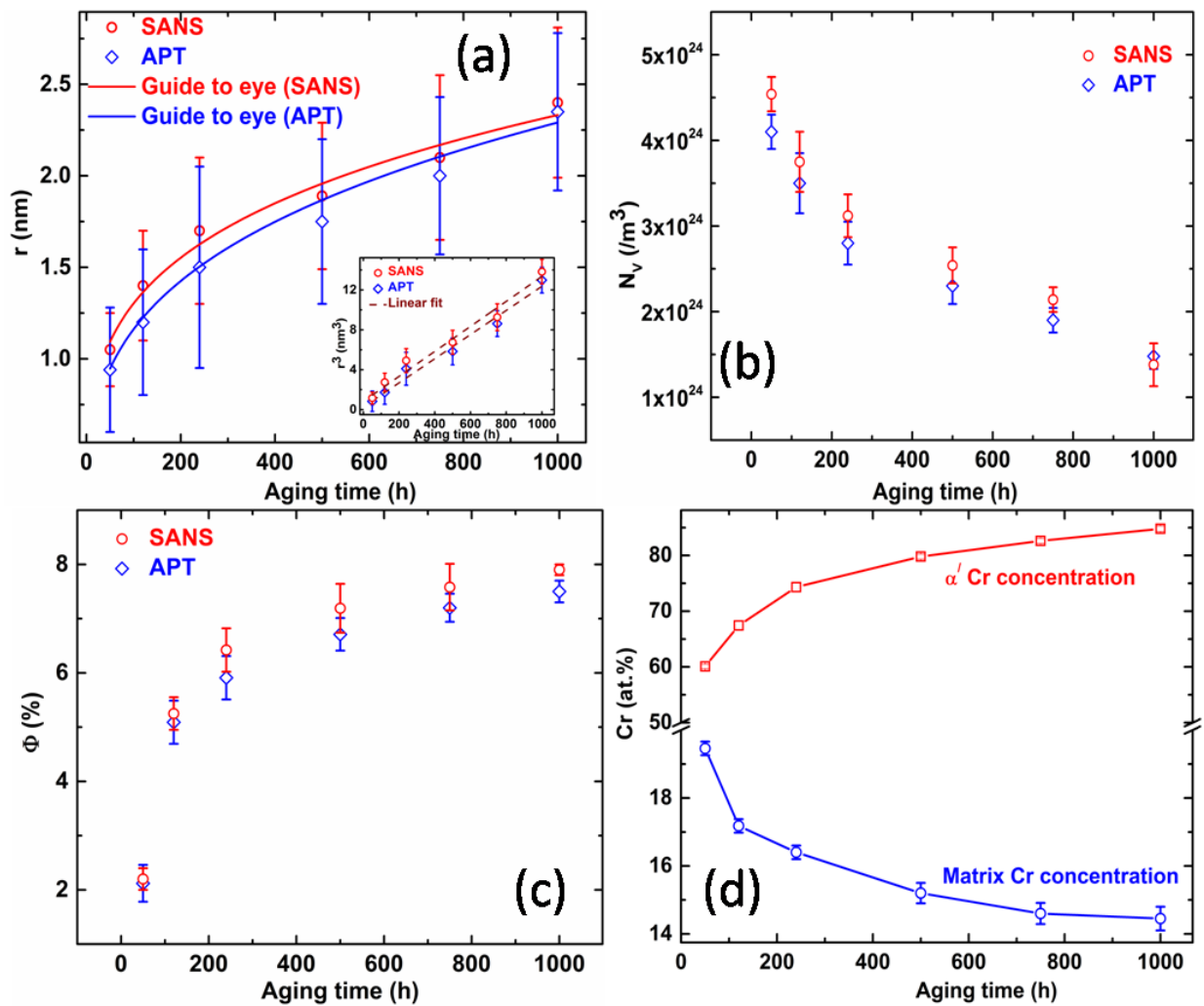

**Fig. 5.** Temporal evolution of: (a) average radius (*r*); while inset represents $r^3$, (b) volume fraction, (c) number density as obtained from complementary APT and SANS techniques for Fe-20 at.% Cr alloy, (d) Cr-concentration for $\alpha'$ and matrix as obtained from APT.

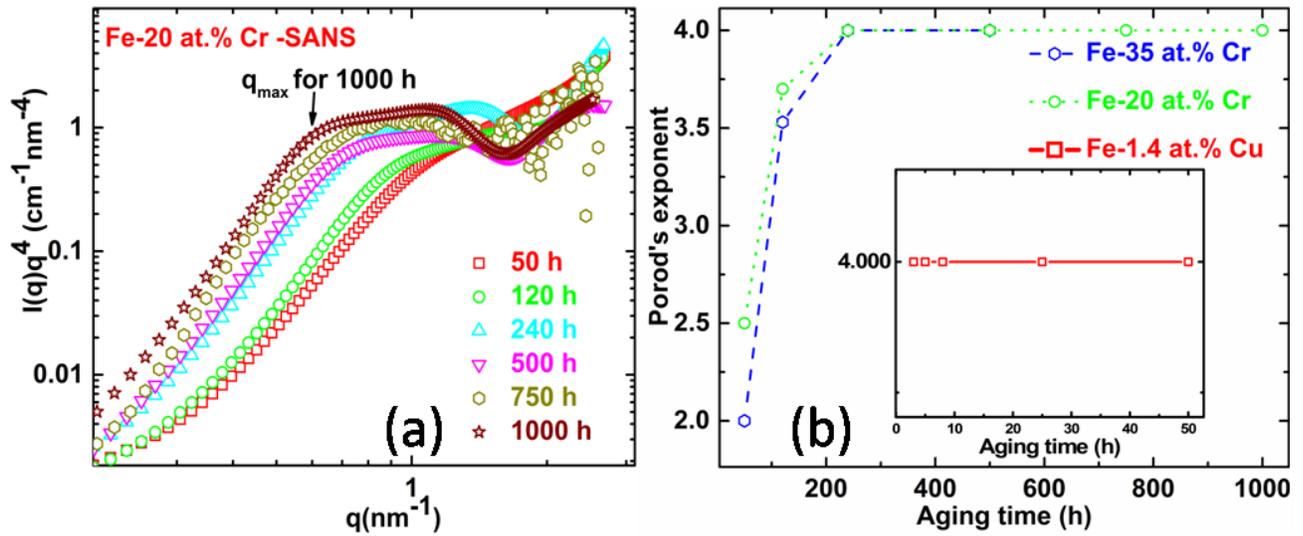

**Fig. 6.** (a) Plateau is observed from 240 h of aging on $I(q)q^4$ vs $q$ plot suggesting Porod's slope of 4 after correlation peak, and (b) temporal evolution of Porod's exponent as a function of aging time as displayed for Fe-20 at.% Cr, Fe-35 at.% Cr and Fe-1.4 at.% Cu alloys respectively, while inset shows magnified image for the Fe-1.4 at.% Cu only.

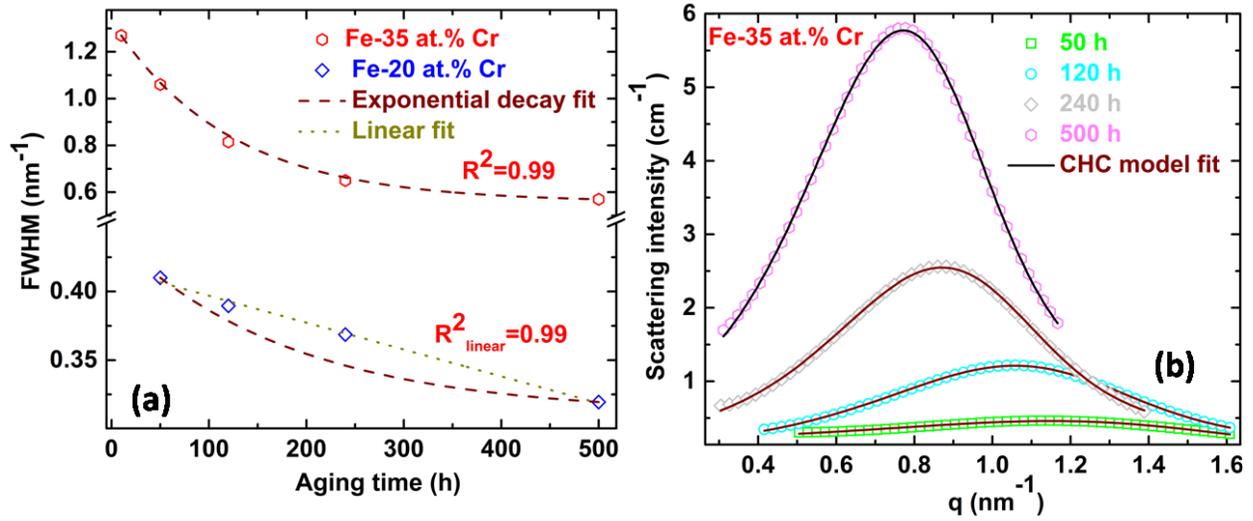

**Fig. 7.** (a) FWHM of the SANS correlation peak as a function of aging time along with exponential decay fit as shown for Fe-20 at.% Cr alloy and Fe-35 at.% Cr alloy, respectively, and (b) CHC model fit of the correlation peaks for Fe-35 at.% Cr alloys.

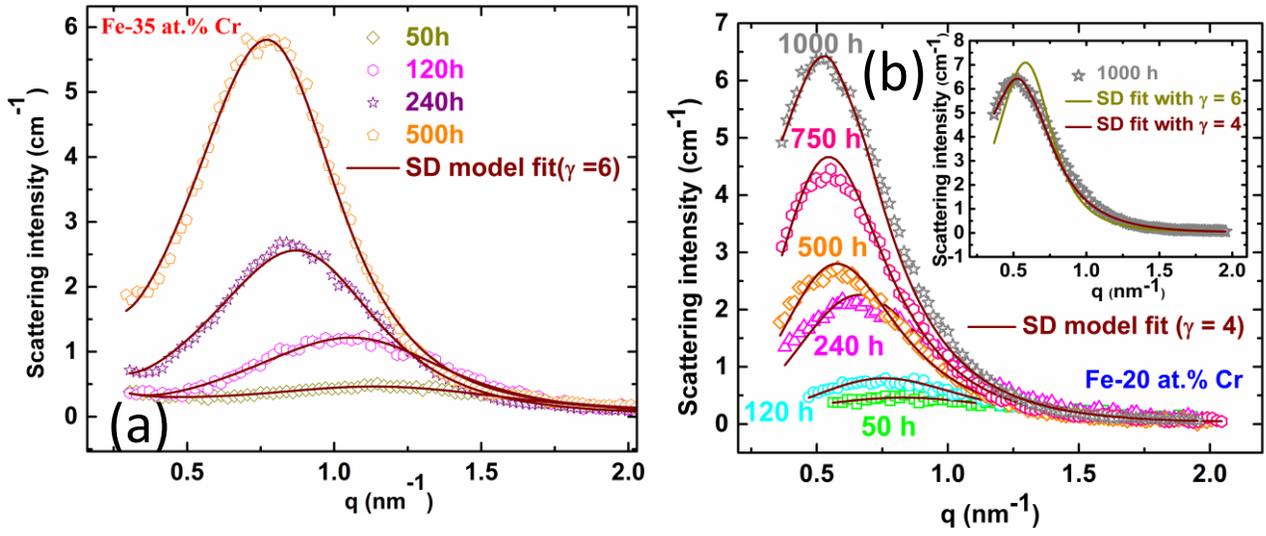

**Fig. 8.** (a) Fitting of correlation peaks with SD model with $\gamma$ value of 6 for Fe-35 at.% Cr alloy aged for 50-500 h, (b) Fitting of correlation peaks with SD model with $\gamma$ value of 4 for Fe-20 at.% Cr alloy aged for 50-1000 h, while inset shows comparison of SD model fitting of 1000 h aged Fe-20 at.% Cr SANS pattern with $\gamma$ value of 4 and 6, respectively.

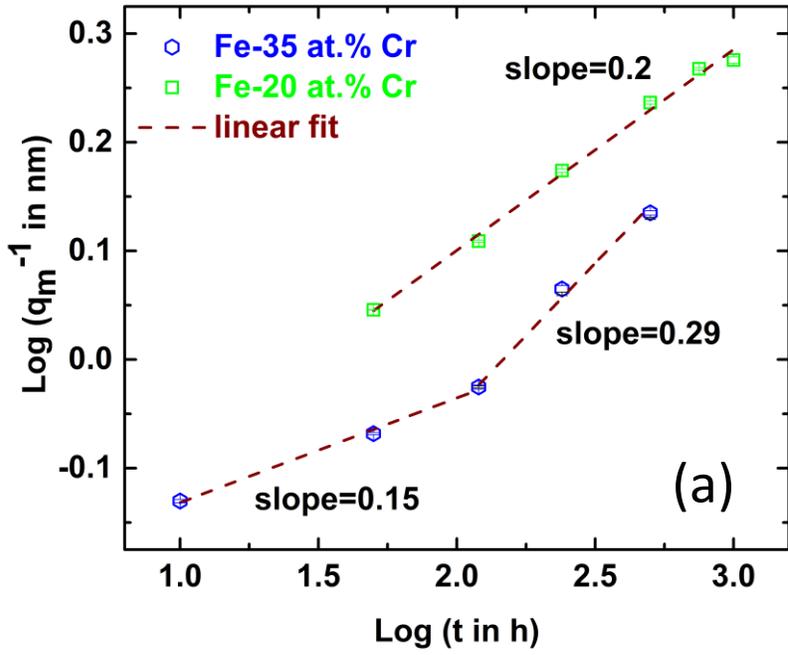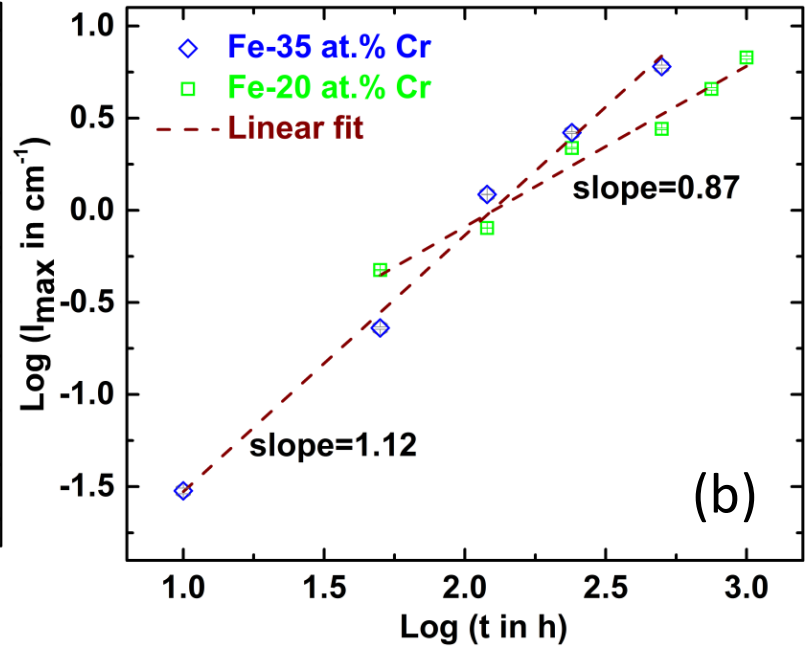

**Fig. 9.** Log-log variations along with linear fits (dashed line) for: (a) $q_m^{-1}$ vs $t$, (b) $I_{max}$ vs $t$.

# List of Tables

## Table 1

**Table 1.** Chemical composition of the solutionized alloys as obtained from WDS-XRF

| Alloys | Fe | Cr | S | P |
|---|---|---|---|---|
| Fe-20 at.% Cr | 80.2 ± 0.1 | 19.8 ± 0.1 | < 0.02 | < 0.02 |
| Fe-35 at.% Cr | 65.1 ± 0.1 | 34.9 ± 0.1 | < 0.03 | < 0.02 |

## Table 2

**Table 2.** CHC model fit parameters for Fe-35 at.% Cr alloy

| Aging time (h) | $2Mk$ (s$^{-1}$) | $A/2k$ (nm$^{-2}$) | $RT/2kc(1-c)$ (nm$^2$) |
|---|---|---|---|
| 50  | 0.48 x 10$^{-6}$ | -2.08 | 0.18 |
| 120 | 1.69 x 10$^{-6}$ | -1.54 | 0.39 |
| 240 | 1.84 x 10$^{-6}$ | -1.06 | 0.52 |
| 500 | 1.93 x 10$^{-6}$ | -0.82 | 0.98 |

**Table 3.** a' and a" as available in literature

| Alloy | Aging temperature (°C) | a' | a" | References |
|---|---|---|---|---|
| Fe-35Cr | 500 | 0.16 | 0.64 | (Xu et al., 2016) |
| Fe-40Cr | 450 | 0.12 | 0.55 | |
| | 500 | 0.16 | 0.82 | |
| Fe-32Cr | 450 | 0.13 | 0.48 | (Katano & Iizumi, 1983) |
| | 500 | 0.18 | 0.62 | |
| Fe-24Cr | 500 | 0.20 | 0.36 | |
| Fe-35Cr | 500 | 0.11 | 0.86 | (Hörnqvist et al., 2015) |
| Fe-32Cr | 500 (850 sol.) | 0.12 | 1.0 | (LaSalle & Schwartz, 1986) |
| | 500 (1200 sol.) | 0.2 | 0.6 | |